# Nonlinear Fokker-Planck equation for nonlocal and nonconservative model Hamiltonians: symplectic integration and application to metastable systems

E. Klotins

*Institute of Solid State Physics, Riga, Latvia*

Kinetics of metastable systems modeled by Hamiltonians containing nonlocal and nonconservative terms is reproduced by the Fokker-Planck and imaginary time Schrödinger equation scheme with subsequent symplectic integration. Example solutions for ergodicity breaking in ferroelectrics reassure the H-theorem of global stability [M. Shiino, Phys. Rev. A, Vol 36, pp. 2393-2411 (1987)] and reproduce spatially extended polarization response in time-dependent external fields.

*Subject class: Statistical Mechanics.*

1. Introduction

Dynamics of metastable systems in general and the polarization response of ferroelectric like systems in particular is an active field of study based on model Hamiltonians emerging as a result of filtering out microscopic degrees of freedom [1]. Langevin and Fokker-Planck equations naturally appear in the calculations and comprise regular terms generated by the model Hamiltonian as well as these contributed from the degrees of freedom which are not essential to the macroscopic variables of interest and removed by the coarse graining. In case of nonlocal and nonconservative model Hamiltonians the Fokker-Planck equations become more complex and its solution crucially depends on the capacity of mathematical technique. In this context the Wentzel-Kramer-Brilluin (WKB) analysis based on mapping between Fokker-Planck and imaginary time Schrödinger equation has received a renewed attention [2] including recent application to quartic model Hamiltonians and linear Fokker-Planck approach reproducing, in some extent, dynamic hysteresis [3].
The subject of present paper is nonlinear Langevin – Fokker-Planck – imaginary time Schrödinger equation technique emerging for systems exhibiting ergodicity breaking. The corresponding experimental situation is addressed to critical phenomena under alternate source field and additive noise and exemplified by polarization response in ferroelectrics. The related fields are stochastic resonance [4,5], and noise activated sensors [6], to name only a few.

General framework of the analysis is given in Sect.2 with special emphasis on the mapping between Fokker-Planck and imaginary time Schrödinger equation, the symplectic integration of linear Fokker-Plank equation, and its comparison between well accepted eigenfunction approach [7] and ab initio results [8]. Ergodicity breaking in the model of globally coupled anharmonic oscillators reproducing fine details of polarization switching is given in Sect.3. Polarization response in systems modeled by locally coupled anharmonic oscillators is reproduced in Sect.4 recovering the size effect on remnant polarization at the initial stage of domain switching. The physical background accepted in the Langevin - Fokker-Planck scheme as well as key problems for further developments are analyzed in Sect.5. Final conclusions are presented in Sect.6.

## 2. Model Hamiltonians: key relations

The Fokker-Planck and imaginary time Schrödinger equation scheme starts with model Hamiltonians. A prototypic example accentuated in this section, is the Ginzburg-Landau type model Hamiltonian reproducing dynamic hysteresis of electric polarization. Specific features of ergodicity breaking and bifurcation of relaxation time appear as a natural extension of this scheme.

The dimensionless Fokker-Planck equation for probability density of polarization $\rho(P,t)$ reads as

$$\frac{\partial \rho(P,t)}{\partial t} = \frac{\partial}{\partial P}\left(\frac{\delta U}{\delta P}\rho(P,t)\right) + \Theta \frac{\partial^2 \rho(P,t)}{\partial P^2} \qquad (1)$$

Here

$$U = -P^2/2 + P^4/4 + (\nabla P)^2/2 - \lambda(t)P, \qquad (2)$$

is the quartic dimensionless energy functional of the Hamiltonian $H = \int U(\mathbf{P})dV$ and the integration over volume is ignored because it is not essential for further approach. The diffusion coefficient $\Theta$ emerges from the microscopic degrees of freedom which are not essential to the macroscopic variables of interest and removed by the coarse graining. The factor $\lambda(t)$ denotes time dependent driving (source) field, and the gradient term $(\nabla P)^2$ specify weak nonlocality. In this section the gradient term in Eq.(2) is omitted for mathematical convenience and the concept is to transform Eqs.(1,2) in imaginary time Schrödinger equation for the auxiliary function $G(P,t)$ introduced by the standard WKB ansatz [2]

$$\rho(P,t) = \exp[F(P)]G(P,t) \qquad (3)$$

The imaginary time Schrödinger equation reads as

$$\frac{\partial G(P,t)}{\partial t} = \left[\Theta \frac{\partial^2}{\partial P^2} + V(P)\right] G(P,t) \tag{4}$$

and the potential operator $V(P)$ is given by

$$V(P) = \left[-\frac{1}{4\Theta}[U'(P)]^2 + \frac{1}{2}U''(P)\right] \tag{5}$$

The auxiliary function $G(P,t)$ Eq.(4) unfolds polarization kinetics through the first moment of probability density $\rho(P,t)$. The survey of computations includes analytical solution of an ordinary differential equation for $F(P)$ in Eq.(4) canceling the first derivative of auxiliary function in Eq.(4) and simultaneously determining the WKB ansatz as

$$\rho(P,t) = \exp[-U(P)/2\Theta] G(P,t) \tag{6}$$

The mapping between Eqs.(1,4) is quite general and applicable also for nonlocal energy functionals. The analytical and quite exact part of computations is completed by recurrence relation for the auxiliary function valid for a small time slice $\Delta t$

$$G(P, t+\Delta t) = \exp\left[\Delta t\left(\Theta \frac{\partial^2}{\partial P^2} + V(P)\right)\right] G(P,t) \tag{7}$$

The symplectic integrator for Eqs.(6,7) reads as

$$\left(1 - \frac{\Theta \Delta t}{2} \frac{\partial^2}{\partial P^2}\right) G(P, t+\Delta t)$$
$$= \left\{\exp\left[\frac{\Delta t}{2}V + \frac{\Delta t^3}{48}(\nabla V)^2\right]\left(1 + \frac{\Theta \Delta t}{2} \frac{\partial^2}{\partial P^2}\right)\exp\left[\frac{\Delta t}{2}V + \frac{\Delta t^3}{48}(\nabla V)^2\right]\right\} G(P,t) \tag{8}$$

Here the potential operator $V$ Eq.(5) is given with time argument $t := t + \frac{\Delta t}{2}$ [9,10]. Advantages of the symplectic integrator Eq.(8) are norm conservation, perfect stability, and suitability for non-conservative and nonlocal (depending on the product of fields at different points) energy functionals. The simplest example of aforementioned approach is dynamic hysteresis mimicking the combined effect of periodic source field and noise in a metastable system exhibiting, in the absence of external source, a unique ground state. Under external source field, if its rate is lesser than another intrinsic time-scale, the equilibrium holds at the instantaneous value of the source field. In general, the metastable states cannot follow the external source field and exhibit itself as the dynamic hysteresis. A stationary solution for a system specified by quartic potential is given in [7] and its application to dynamic hysteresis is attempted in [3]. The results shown in Fig.1 confirm, for a representative set of parameter values, that the agreement between [7] and the test solution results preceded by Eq.(6) for the dynamic hysteresis [3] is fairly good and reassures accuracy of the approach Eqs.(1-7). In more detail the solution [7] is semiadiabatic in the sense that only the first non-zero eigenvalue contributes in the dynamics. Otherwise, no explicit restrictions for time scales are defined for the Fokker-Planck imaginary time Schrödinger approach Eqs. (1-7).

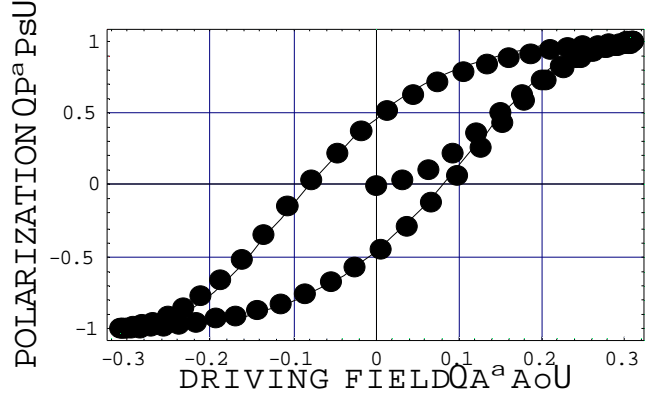

Fig.1 Dynamic hysteresis under harmonic field: semiadiabatic approach [7] (line) and symplectic integration (dots). Parameters of the problem: amplitude of the source field $\lambda = 0.309\lambda_0$, $\lambda(t) = \lambda \sin(\Omega t)$, low driving frequency $\Omega = 10^{-3}$, diffusion constant $\Theta = 1/20$, and $\lambda_0 = 2/\sqrt{27}$ is the static coercive field.

Another example concern the $(\Psi^2)^3$ energy functional $U(P_1) = \alpha_1 P_1^2 + \alpha_{11} P_1^4 + \alpha_{111} P_1^6$. For $PbTiO_3$ [11] the expansion coefficients are: Curie-Weiss constant $1.5\,10^5\,^0C$, transition temperature $T_C = 492.2$ (°C), $\alpha_1 = 61 \cdot 10^5$ (m/F) at $T_C$, $\alpha_{11} = -9.235 \cdot 10^7$ (m$^5$/(C$^2$F), $\alpha_{111} = 3.469 \cdot 10^8$ (m$^9$/(C$^4$F).

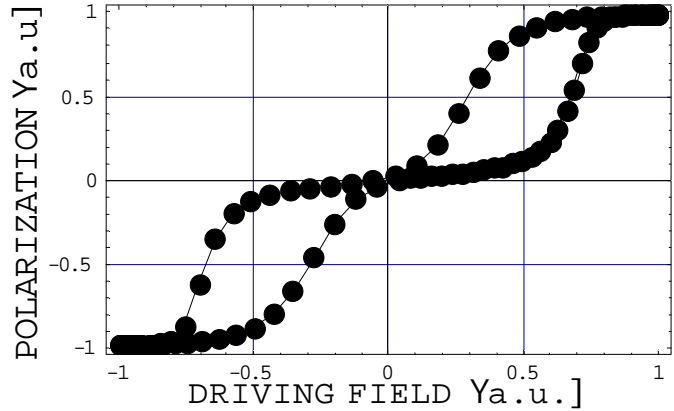

Fig.2 Dynamic hysteresis plot for $PbTiO_3$ at temperature $T = 1.005\,T_C$. The source field is given in units of a $0.15$ maximum value at which the energy landscape transforms in a single wall one. The rest of parameters are as follows: diffusion constant $1/100$, frequency $10^{-3}$. The polarization is normalized by the spontaneous one.

The third example is focused on a connection between the abovementioned phenomenological approach and the first principles calculations [8] in which the nonlinear dielectric and piezoelectric response of tetragonal $PbTiO_3$ is modeled for fields along the $z-$direction. The corresponding fitting with Landau-Devonshire expansion reads as (in denominations of [8])

$$F(\eta_1, \eta_3, P_z) = \frac{1}{2}c_{11}(2\eta_1^2 + \eta_3^2) + c_{12}(2\eta_1\eta_3 + \eta_1^2) + A_2 P_z^2 + A_4 P_z^4 + A_6 P_z^6 + 2B_{1yy}\eta_1 P_z^2 + B_{1zz}\eta_3 P_z^2 \quad (9)$$

Here $\eta_1 = \eta_{xx} = \eta_{yy}$ and $\eta_3 = \eta_{zz}$ are components of strain tensor, and the expansion is truncated to sixth order in polarization $P_z$, and the first order in elastic and polarization-strain coupling. Numerical values of the parameters for $PbTiO_3$ are as follows: $A_2 = -0.003$, $A_4 = 0.005$, $A_6 = 0.004$ (in the units of $Ha$ and $C/m^2$). The components of elastic tensor and the coefficients of polarization-strain coupling are $c_{11} = 4.374$, $c_{12} = 1.326$, $B_{1zz} = -1.99$ and $B_{1yy} = -0.049$, correspondingly. Clamped-strain response is obtained by zeroing the variations of Eq.(9) with respect to the components of strain tensor and putting back in Eq.(9). As a result the expansion coefficient at $P^4$ in Eq.(9) renormalizes as follows[8]

$$F(P_z) = A_2 P_z^2 + \left( A_4 + \frac{2c_{12}B_{1zz}B_{1yy} - c_{11}B_{1yy}^2 - \frac{1}{2}(c_{11} + c_{12})B_{1zz}^2}{(c_{11} + 2c_{12})(c_{11} - c_{12})} \right) P_z^4 + A_6 P_z^6 \quad (10)$$

Since the first expansion coefficient $A_2 < 0$ the energy landscape exhibit two local minima and its evolution with source field yields static hysteresis loop shown in Fig.3. Dynamic hysteresis appears as interplay between internal dynamics of the system and the period of alternate source field. Formally this interplay is mimicked by the Langevin – Fokker-Planck scheme quite similarly to this in Eqs.(1-8). It starts with variation of Eq.(10) and yields Langevin equation

$$\frac{\partial P_z}{\partial t} = -\gamma \frac{\delta[F(P_z) - P_z\lambda(t)]}{\delta P_z} + \eta(t) \quad (11)$$

comprising three fitting parameters: prefactor (kinetic coefficient) $\gamma$, period of source field, and the noise strength $\eta(t)$. At appropriate choice of these parameters, the kinetics is nonadiabatic and the dynamic hysteresis plot fits fairly well with the static one.

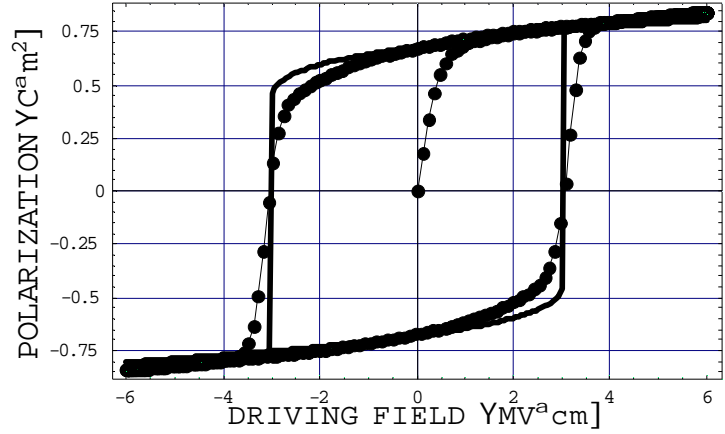

Fig.3 Comparision of first principle clamped-strain static hysteresis plot (bold line)[8] and the polarization response on periodic source (dots).At high frequency source field ($\Omega = 10^{-2}$) the dynamic hysteresis plot approaches to the first principle one and the coercive field resembles the static one, as expected. The rest parameters are: diffusion coefficient $1/1000$, kinetic coefficient $\gamma = 10$.

Model energy functional Eq. (9) [8] and its extensions is an example of successful coarse graining at which the polarization of electrons and nonlinear lattice polarization are abandoned for convenience of calculations. However, in spite of considerable efforts [12,13] to determinate fine details of the emergence of noise the available application-grade practice is to precede coarse graining of already coarse grained models obeying the idea that physics on different scales decouple and critical phenomena can be described by finite number of macroscopic variables.

3. Ergodicity breaking in the model of globally coupled anharmonic oscillators

The key signature of ergodicity breaking is the existence of some operator, namely order parameter, whose ground-state expectation value undergoes bifurcation and divergence of the relaxation time.

A representative example for this kind of phenomena is the system modeled by coupled overdamped anharmonic oscillators [14] interacting by way of attractive linear coupling. A set of Langevin equations for the order parameter $P(t)$ reads as

$$\frac{\partial P_i}{\partial t} = -\frac{\partial F(P_i)}{\partial P_i} + \sum_{k=1}^{N} \frac{\varepsilon}{N}(P_k - P_i) + \eta_i(t) \qquad (12)$$

here the stochastic force terms $\eta_i(t)$ determines the simplest bath customary specified by statistically independent white noise with correlation function $\langle \eta_i(t)\eta_j(t')\rangle = \delta_{ij}\delta(t-t')$ and the factor $\varepsilon > 0$ denotes the strength of attractive mean-field type coupling (at $N=1$ Eq.(12) reduces to Eq.(2)). At the thermodynamic $N \to \infty$ limit the averages of $P_k$ in Eq. (12) can be assumed to behave in a deterministic way, namely, $\lim_{N\to\infty}\left(\frac{1}{N}\sum_{k=1}^{N}P_k(t)\right) = \overline{P}(t)$ and the corresponding Fokker-Planck equation concern probability density for each $P_i$ which originates from various realizations of white noise

$$\frac{\partial \rho}{\partial t} = \sum_{i=1}^{N}\left[-\frac{\partial}{\partial P_i}\left[-\frac{\partial F}{\partial P_i} + \frac{\varepsilon}{N}\sum_{k=1}^{N}P_k - \frac{\varepsilon}{N}\sum_{k=1}^{N}P_i\right]\rho + \Theta\frac{\partial^2 \rho}{\partial P_i^2}\right] \tag{13}$$

Recognizing that $\frac{\varepsilon}{N}\sum_{k=1}^{N}P_k(t) = \varepsilon\overline{P}(t)$ and each $i$-th entity concerns a coarse-grained block described by equal kinetics Eq.(12) the Eq.(13) reduces to

$$\dot{\rho}(P,t) = \frac{\partial}{\partial P}\left[U'(P,t) + \Theta\frac{\partial}{\partial P}\right]\rho(P,t) \tag{14}$$

For quartic energy functionals [14] $U(P,t) = -\frac{P^2}{2} + \frac{P^4}{4} - P\lambda(t) + \frac{\varepsilon}{2}\left[P - \overline{P}(t)\right]^2$ and $\overline{P}(t)$ is the expectation value of order parameter that emerges a substantional nonlinearity in the Fokker-Planck equation Eq.(14) managed, as we shall show hereafter, by symplectic integration. Finally, the Eq.(14) is a intermediate mesoscopic scale description comprising stochastic elements due the effect of coarse-graining. For mathematical convenience the local and nonlinear terms in Eq. (14) are represented by $U_1(P,t) = -\frac{P^2}{2} + \frac{P^4}{4} - P\lambda(t)$ and $U_2(P,t) = \frac{\varepsilon}{2}\left[P - \overline{P}(t)\right]^2$, correspondingly, resulting in the stationary solution (assumed as the initial condition for further calculations) as

$$\rho(P,0) = \frac{\exp\left[-\frac{U_1(P,0)+U_2(P,0)}{\Theta}\right]}{\int \exp\left[-\frac{U_1(P,0)+U_2(P,0)}{\Theta}\right]dP} \tag{15}$$

Here the denominator provides normalization of the probability distribution $\int \rho(P,0)dP = 1$ and the stationary (initial) value of the first moment of polarization density $\overline{P}(0)$ is found by integrating Eq.(14) over $P$ with $\overline{P}(0)$ as a parameter. In $P, P - \overline{P}(0)$ frame the ergodicity breaking occurs if $\partial(P - \overline{P}(0))/\partial P < 0$ and the exact value of $\overline{P}(0)$ is found at intersection of $P - \overline{P}(0)$ plot with the $P$ axis as shown in Fig.4.

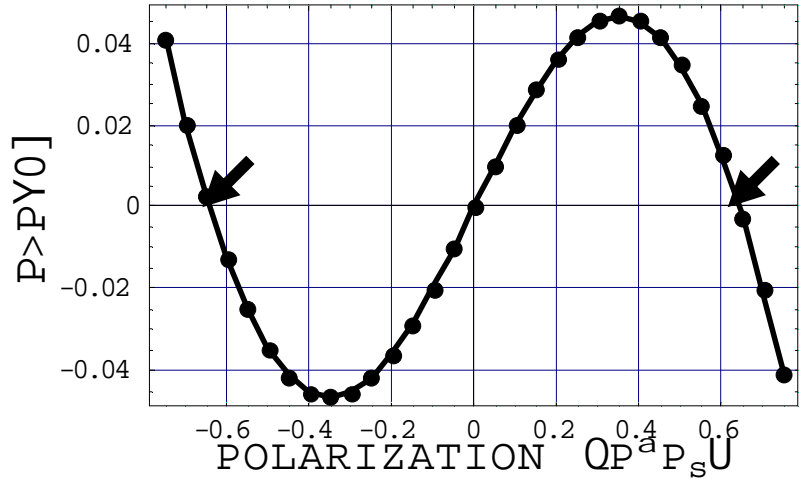

Fig. 4 Schematic plot of $P - \overline{P}(0)$ as a function of polarization $P$ (in $P_s$ units). The stationary polarization $\overline{P}(0)$ is recovered by intersection of $P - \overline{P}(0)$ plot with the $P$ axis exhibiting two stable solutions at $\overline{P} = \pm 0.64$ (marked by arrows) and an unstable solution at $\overline{P} = 0$. Representative parameters of the model: $\varepsilon = 0.07$, $\lambda = 0$, $\Theta = 1/20$.

The bifurcation of ground state critically depends on the diffusion and coupling constants. Another consequence is that the different ground states are indistinguishable unless a symmetry-breaking interaction is introduced with external or internal fields. Similarly, the stationary solution of Eq.(14) for $(\Psi^2)^3$ type model Hamiltonians crucially depends on the temperature and exhibit ground-state bifurcation at $T < T_C$. Otherwise, at $T > T_C$ zero ground-state prevails except a small symmetry breaking voltage is applied. This stationary state is supported by the Boltzmann's H-theorem (which ensures the existence of a uniquely determined long-time probability distribution $\rho_\infty(P,t)$) and, in case of nonlinearity [14], stating that (at overcritical interaction constant) the system always reaches global stability in the sense that any time dependent solution of Eq.(14) lying far from equilibrium must be attracted by either one of those stationary solutions without any possibility of runaway behavior or limit cycle type oscillations. The source pulse initiated temporal behavior is modeled taking advance from the Fokker-Planck – imaginary time Schrödinger scheme starting with the ansatz

$$\rho(P,t) = \exp\left[-\frac{U_1(P,t)+U_2(P,t)}{2\Theta}\right]G(P,t) \tag{16}$$

The auxiliary function $G(P,t)$ given by nonlinear imaginary time Schrödinger equation

$$\dot{G}(P,t) = \left[\Theta\frac{\partial^2}{\partial P^2} + V_1(P,t) + V_2(\overline{P}(t),P,t)\right]G(P,t) \tag{17}$$

made up of both the linear $V_1(P,t)$ and the nonlinear $V_2(\overline{P}(t),P,t)$ terms in the potential operator. Here

$$V_1(P,t) = -\frac{1}{4\Theta}\left[\frac{\partial U_1(P,t)}{\partial P}\right]^2 + \frac{1}{2}\frac{\partial^2 U_1(P,t)}{\partial P^2} + \frac{1}{2\Theta}\left[\frac{\partial U_1(P,t)}{\partial t}\right] \tag{18}$$

and

$$V_2(\overline{P}(t),P,t) = \\ \frac{1}{2\Theta}\frac{\partial U_2(P,t)}{\partial t} - \frac{1}{2\Theta}\frac{\partial U_1(P,t)}{\partial P}\frac{\partial U_2(P,t)}{\partial P} - \frac{1}{4\Theta}\left(\frac{\partial U_2(P,t)}{\partial P}\right)^2 + \frac{1}{2}\frac{\partial^2 U_2(P,t)}{\partial P^2} \tag{19}$$

Example solution for temporal polarization response of a system with initially positive remnant polarization affected by a negative sow tooth shaped source pulse is demonstrated in Fig.5. The source pulse is specified by $-0.027$ (a.u.) amplitude (corresponding $\sim 0.07$ of the thermodynamic coercive field in physical units) and its length varying between $320 \leq T \leq 1580$ a.u. What is anticipated after the source pulse turns to zero is approaching the expectation value $\overline{P}$ to a remnant polarization, either $P_r$ or $-P_r$. The sign of the expectation value $\overline{P}$ at the time instant $T$ at which the source turns to zero is crucial, namely, at $\overline{P}(t=T) > 0$ the remnant polarization approaches to $\overline{P}(t=\infty) \to \overline{P}_r$, and $\overline{P}(t=\infty) \to -\overline{P}_r$ otherwise as it follows from the H-theorem of global stability [14]. This behavior is confirmed in Fig.5 with $\overline{P}(T) = 0$ as the point splitting the $\overline{P}$ - space in two domains of attraction for $P_r$ and $-P_r$. However, time propagation of the system at $0 < t < T$ within which the source field is nonzero goes beyond the H-theorem of global stability [14] and is revealed (for this representative set of parameters) by the aforementioned scheme. As shown in Fig.5 all pulses of length exceeding $T \geq 710$ are obviously overcritical and belong to the $-P_r$ domain of attraction. Unlike this, the pulses of length $T < 640$ are undercritical and belong to the $P_r$ domain of attraction.

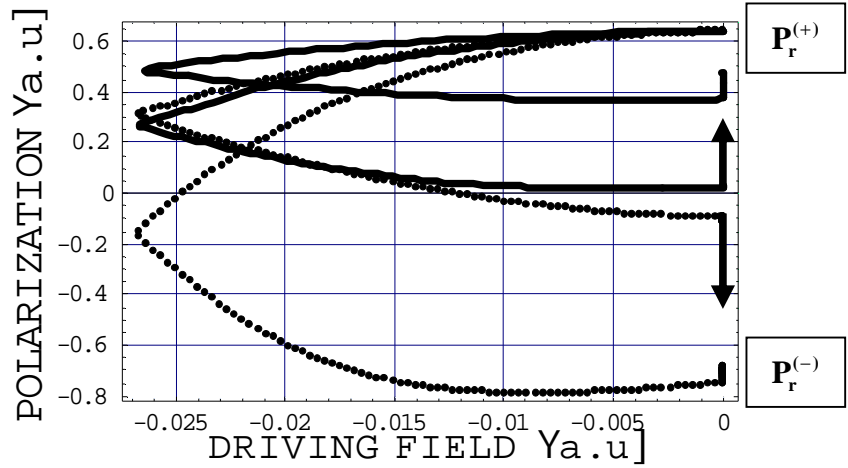

Fig. 5 Time propagation of polarization switched from $\overline{P}_r(t=0)=+0.64$ stationary (initial) state to the final state as initiated by $-0.027$ sow teeth shaped pulses of $320, 640, 710$, and $1580$ (a.u.) length. The plots (lines) illustrate the effect of $T=320$ and $T=640$ pulses being to short for polarization switching. So, in spite of nonzero source, the system remains in the $P_r^{(+)}$ domain of attraction. Otherwise, starting from pulse length $T=710$ (dots) the system enters in the $P_r^{(-)}$ domain of attraction.

What is missing in this analysis is the spatial extension lost in the case of globally coupling. Nevertheless, the aforementioned mathematical technique, namely, the implementation of recursion-specific first moment $\overline{P}(t_i)$ Eq.(15) has essential consequences for spatially extended problem emerged by locally coupled model in which each anharmonic oscillator is coupled which its first neighbors.

### 4. Polarization response in the model of locally coupled anharmonic oscillators

Spatial dependence of polarization field, disappearing in the model of globally coupled anharmonic oscillators Eqs.(12) may be restored ad hoc considering the model of first neighbors coupling. This approach assumes that (i) the system consist of finite number of (microscopically large) blocks modeled by Ginzburg-Landay energy functional $\Phi_i = -\frac{1}{2}P_i^2 + \frac{1}{4}P_i^4 - \lambda(t)P_i$ (here we consider the $T<T_C$ case and the sixth order term is irrelevant) and (ii) the first neighbor interaction between (macroscopically small) blocks holds so addressing the problem to ensemble of interacting blocks. Going around the microscopic interpretation of the strength of interaction and the correlation length, the problem is formulated by the model Hamiltonian

$$H \equiv \sum_i^N \left\{ \Phi_i + \frac{\varepsilon}{2} \left( (\overline{P}_{i+1}(t) - P_i)^2 + (\overline{P}_{i-1}(t) - P_i)^2 \right) \right\} \quad (20)$$

Here the expectation values $\overline{P}_k$ are unknown quantities and are evaluated selfconsistently afterward. Kinetic equations derived from Eq.(20)

$$\frac{\partial P_i}{\partial t} = -\frac{\partial \Phi_i}{\partial P_i} + \varepsilon \left( \overline{P}_{i+1}(t) - 2P_i + \overline{P}_{i-1}(t) \right) \quad (21)$$

readdress the problem to a set of Fokker-Planck equations for probability distribution

$$\dot{\rho}(P_i, t) = -\frac{\partial}{\partial P_i} \left[ -\frac{\partial \Phi_i}{\partial P_i} \rho(P_i, t) + \varepsilon \left( \overline{P}_{i+1}(t) - 2P_i + \overline{P}_{i-1}(t) \right) \rho(P_i, t) \right] + \Theta_i \frac{\partial^2}{\partial P_i^2} \rho(P_i, t) \quad (22)$$

In stationary case

$$\rho(P_i) \left( 2\varepsilon + \frac{\partial \Phi_i}{\partial P_i^2} \right) + \left( \varepsilon (\overline{P}_{i-1} - 2P_i + \overline{P}_{i+1}) + \frac{\partial \Phi_i}{\partial P_i} \right) \frac{\partial \rho(P_i)}{\partial P_i} + \Theta \frac{\partial^2 \rho(P_i)}{\partial P_i^2} = 0 \quad (23)$$

and the stationary probability distribution yields

$$\rho(P_i) = C \exp\left[ \frac{-\Phi(P_i) + \varepsilon P_i (\overline{P}_{i-1} - P_i + \overline{P}_{i+1})}{\Theta_i} \right] \quad (24)$$

Here $C$ is normalization constant, and $\overline{P}_0 = 0$, $\overline{P}_{i_{max}+1} = 0$ are zero boundary conditions. Implementing normalization of the probability distribution as well as the first moment $\overline{P} = \int_{-\infty}^{\infty} P \rho dP$ the selfconsistency condition for $\overline{P}_i$ is given by

$$\frac{\int_{P_{min}}^{P_{max}} P_i \exp\left[ \frac{-\Phi(P_i, 0) + \varepsilon P_i (\overline{P}_{i-1} - P_i + \overline{P}_{i+1})}{\Theta_i} \right] dP_i}{\int_{P_{min}}^{P_{max}} \exp\left[ \frac{-\Phi(P_i, 0) + \varepsilon P_i (\overline{P}_{i-1} - P_i + \overline{P}_{i+1})}{\Theta_i} \right] dP_i} - \overline{P}_i = 0 \quad (25)$$

For a set of starting values obtained, for example, from a static approach [15] Eq. (25) gives the stationary solution of Cauchy problem. Figure 6 demonstrates example solutions of Eq.(25) for zero boundary conditions, $\overline{P} = \pm 1$ starting values, and various coupling between blocks as a parameter.

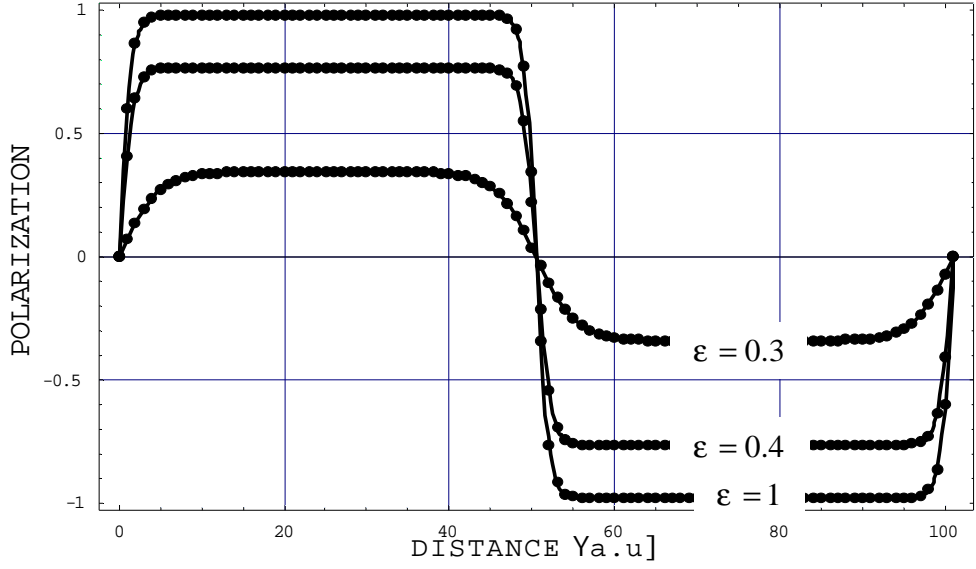

Fig. 6 Example solution for $180^0$ domains in a 1D region with zero boundary conditions and $0.3, 0.4, 1$ coupling constants. The difference between the spontaneous and remnant polarization disappears at enlarging the coupling constant as illustrated by the $\varepsilon = 1$ plot.

Nonstationary solution of Eq.(23) starts with the ansatz

$$\rho(P_i,t) = \exp[-F(P_i,t)]G(P_i,t) \tag{26}$$

Inserting the ansatz Eq. (26) in Fokker-Planck equations Eqs. (22) and zeroing the coefficients at $\partial G(P_i,t)/\partial P_i$ yields the relation for exponential factor in Eq.(26)

$$F_i = \frac{\varepsilon P_i(\overline{P}_{i-1} - P_i + \overline{P}_{i+1}) - \Phi(P_i)}{2\Theta} \tag{27}$$

Substitution of Eq.(27) in Eqs.(26,22) gives the imaginary time Schrödinger equation for the auxiliary function

$$\dot{G}(P_i,t) = [T[i] + V_1[i] + V_2[i] + K[i]]G(P_i,t) \tag{28}$$

Over a spatial mesh $i \in [1, i_{max}]$ the kinetic operator $T[i]$ is given by

$$T[i] = \Theta \frac{\partial^2}{\partial P_i^2} \tag{29}$$

Here the linear part $V_1[i]$ and the nonlinear part $V_2[i]$ of potential operators in Eq.(28) are given by

$$V_1[i] = -\frac{1}{4\Theta_i}\left(\frac{\partial \Phi_i}{\partial P_i}\right)^2 + \frac{1}{2}\left(\frac{\partial^2 \Phi_i}{\partial P_i^2}\right) \tag{30}$$

$$V_2[i] = \frac{\varepsilon\left(4\Theta_i - (2P_i - \overline{P}_{i-1}(t) - \overline{P}_{i+1}(t))\left(2P_i - \varepsilon(\overline{P}_{i-1}(t) + \overline{P}_{i+1}(t)) + 2\frac{\partial \Phi}{\partial P_i}\right)\right)}{4\Theta_i} \quad (31)$$

and the correction to the potential operators generated by explicit time dependence of the energy functional yields as

$$K[i] = \frac{-\varepsilon P_i \dot{\overline{P}}_{i-1}(t) - \varepsilon P_i \dot{\overline{P}}_{i+1}(t) + \dot{\Phi}_i}{2\Theta} \quad (32)$$

Aforementioned analytical calculations are followed by the numerical part comprising the solution of Eq.(28) and evaluation of the merit $M(Q) = \int P\rho(P,Q,t)dP - (\overline{P}(0) + Q\Delta t)$ (transformed in analytical function of $Q_i$ by quadratic interpolation). This trick generates a set of coupled algebraic equations $M(Q_i) = 0$ for expansion coefficients $Q_i$ so returning the density distributions by Eq.(26) over spatial mesh in every time slice.

Preliminary results of the domain switching are shown in Fig.7 for a couple of $180^0$ domains Fig.5. The switching is initiated by an enlarging (negative) source field that reduces the value of polarization in both domains and (at the time instant exposed in Fig.6). It must be emphasized that the difference $\overline{P}(t) - \overline{P}(t)$ being negative for any spatial coordinate prevails at the boundaries and at the domain wall in accord with recent estimates [16].

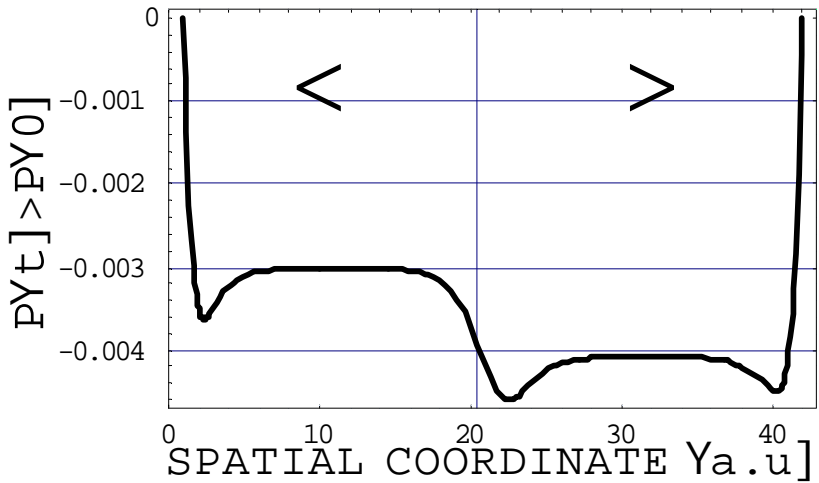

Fig.7 Initial stage of domain switching as initiated at the boundaries and the domain wall. Representative parameters of the problem: first neighbor coupling constant $\varepsilon = 0.035$, diffusion constant $\Theta = 0.05$.

For a rich scale of periodic and spatio-temporal behavior the vector nature of order parameter is crucial and formally contributes in all stages of calculations. Therefore it is reasonable to check validity of the aforementioned scheme taking use of the $(\Psi^2)^3$ type Landau – Ginzburg model [17]. Unlike the the quartic energy functional Eq.(2) the corresponding ionic part of [17,11] reads as follows

$$U(\mathbf{P}) =$$
$$\alpha_1(P_1^2 + P_2^2 + P_3^2) + \alpha_{11}(P_1^2 + P_2^2 + P_3^2)^2 + \alpha_{12}(P_1^2 P_2^2 + P_2^2 P_3^2 + P_1^2 P_3^2) + \alpha_{111}(P_1^6 + P_2^6 + P_3^6) \quad (33)$$
$$+ \alpha_{112}(P_1^4(P_2^2 + P_3^2) + P_2^4(P_3^2 + P_1^2) + P_3^4(P_1^2 + P_2^2)) + \alpha_{123}(P_1^2 P_2^2 P_3^2)$$

Here $\alpha$ are the 3D Landau expansion coefficients. The relevant Fokker-Planck equation for probability density of polarization $\rho(P_1^2, P_2^2, P_3^2, t)$ comes from variation of Eq.(33) with respect to variables $P_1^2, P_2^2, P_3^2$ and unlike Eq. (1) one yields

$$\frac{\partial \rho(\mathbf{P},t)}{\partial t} = \rho(P,t) \sum_{i=1}^{3} \frac{\partial^2 U(\mathbf{P})}{\partial P_i} + \sum_{i=1}^{3} \frac{\partial U(\mathbf{P})}{\partial P_i} \frac{\partial \rho(\mathbf{P},t)}{\partial P_i} + \sum_{i=1}^{3} \Theta_i \frac{\partial^2 \rho(\mathbf{P},t)}{\partial P_i^2} \quad (34)$$

Here the diffusion matrix $\Theta$ is considered as diagonal. Transformation of Eq.(34) into imaginary time Schrödinger equation is irrelevant to the partial structure of Eq.(33) and in stationary state reads as

$$\sum_{i=1}^{3} \left( \Theta_i \frac{\partial^2}{\partial P_i^2} + \left( -\frac{1}{4\Theta_i} \frac{\partial U(\mathbf{P})^2}{\partial P_i} + \frac{1}{2} \frac{\partial^2 U(\mathbf{P})}{\partial P_i^2} \right) \right) G(\mathbf{P}, t=0) = 0 \quad (35)$$

Extension of Eq. (35) for the kinetics of auxiliary function $G(\mathbf{P},t)$ is straightforward and omitted for brevity. The analytical part completes with stationary probability distribution

$$\rho(\mathbf{P}, t=0) = C Exp\left[-\frac{U(\mathbf{P})}{\Theta}\right], \quad (36)$$

the evaluation of WKB ansatz

$$W(\mathbf{P}, t=0) = Exp\left[-\frac{U(\mathbf{P})}{2\Theta}\right] G(\mathbf{P}, t=0) \quad (37)$$

and relation for the initial value of auxiliary function $G(\mathbf{P}, t=0)$. What one needs to start symplectic integration in the 3D case is vector representation of $G(\mathbf{P}, t=0)$ over the polarization mesh. Its structure is different from the zero dimension case Eq.(6) and is illustrated in Fig. 8 for the following representative set of parameters: $P_3 = 0$, $\alpha_1 = -1/2$, $\alpha_{11} = 1/4$, $\alpha_{12} = 1/2$.

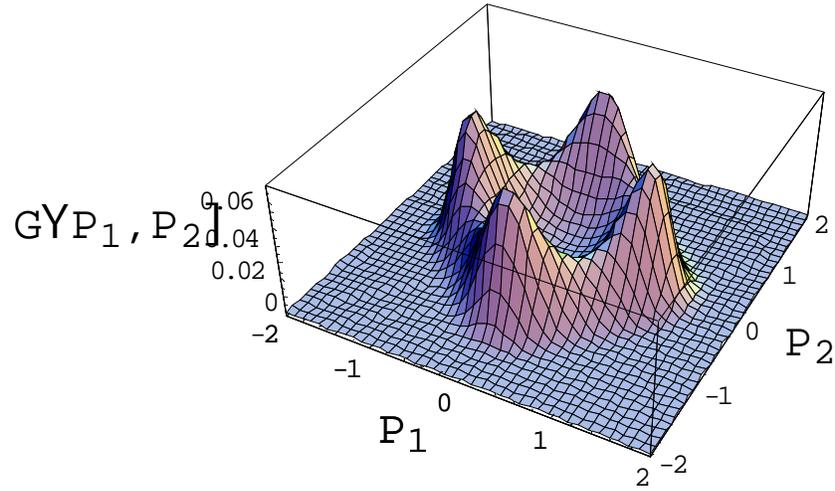

Fig.8 Auxiliary function $G(P_1, P_2, P_3 = 0, t = 0)$ illustrating the way by which two values of polarization $P_1, P_2$ is assigned to each mesh point.

The consequence is that in vector representation any component of auxiliary function $G(P_1, P_2, P_3)$ inherits the impact of all components of the polarization vector. However, the general structure of symplectic integration Eq.(8) remains unchanged and its extension in case of spatial coupling is straightforward.

5. Discussion

The aforementioned approach based on symplectic integration of nonlinear Fokker-Planck equations is a method for calculating properties of metastable systems in time-dependent external fields. In its construction, initially formulated for simplest quartic model Hamiltonians and stochastic terms paralleled with Brownian motion, there is nothing that would exclude arbitrarily nonlinearity, nonlocality and multidimensionality, which imply that this formalism should work in large scale of problems. In this context model Hamiltonians adequate reflecting the microscopic background turns out as the entities of major interest. At the bottom of the hierarchy lays ab initio Hamiltonians and at the macroscopic end, on the other hand, there is nonequilibrium thermodynamics which utilize phenomenological parameters and constitutive equations to yield the dynamics of macro variables. Whereas a systematic hierarchy of model Hamiltonians is out of scope the analysis of where do the (over)simplified model Hamiltonians constituted from a set of anharmonic oscillators (blocks) stand is highly motivated.

Firstly, for globally coupled systems introduced for mathematical convenience, typically each lattice site is connected to all others with the same coupling strength. Physically meaning are microscopically large and macroscopically small objects within which the order parameter is uniform and obey Landau relations. As one can see the mutual effect of coupling and noise reproduces both ordered phase and the ergodicity breaking. Another level of coarse graining emerges as represented by diffusively coupled blocks located at the sites of a lattice with just nearest neighbor coupling. Formally it yields a generalization of usual thermodynamics for spatially inhomogeneous situations, where the order parameter become a coordinate dependent field $\mathbf{P}(\mathbf{x})$ if smoothed over blocks whose center point lies at $\mathbf{x}$ [1].

However, the Langevin – Fokker-Planck and imaginary time Schrödinger equations if written in physical units comprises two parameters, the noise term $\eta(t)$ and the prefactor $\gamma$, the microscopic interpretation of which is still a challenge. These parameters emerge from Langevin equation

$$\frac{\partial P(\mathbf{x},t)}{\partial t} = -\gamma \frac{\delta U[P(\mathbf{x},t)]}{\delta P(\mathbf{x},t)} + \eta(t) \tag{38}$$

The microscopic background, if customary specified by statistical properties of white noise as in Eq.(12), assumes some objects which are different from the system constituent particles and its description is paralleled with the models of Brownian motion. In an advanced approach [12, 13] the noise emerges from model Hamiltonian comprising classic terms for the system and quantum ones for the bath. The problem lies in extension of this quantum Brownian motion to condensed state systems acting as its own bath constituted from the high frequency Fourier modes (or disordered potential) interacting with the order parameter field. Whereas this approach requires still work, reduction to the aforementioned Langevin and Fokker-Planck equations is expected if coupling with the bath is weak and linear.

Another highly motivated problem of critical dynamics is understanding the prefactor $\gamma$ in the kinetic equations Eq. (38). Whereas it formally relies on density functional theories [1] there is, however, no systematic derivation of the dynamic equations governing the time evolution in case of ergodicity breaking. A severe assumption hidden in aforementioned Langevin - Fokker-Planck scheme is that coupling of the prefactor $\gamma$ with the order parameter and other quantities of the theory is not critical. Howevwr, the remaining challenge is to go beyond this mean-field level and climb up a whole hierarchy of first principles based models.

## 6. Conclusions

Noise activated nonadiabatic behavior of metastable systems is investigated and a systematic survey for calculating its behavior under time-dependent external fields presented. Representative examples concern electric hysteresis and polarization switching in presence of ferroelectric phase instability and demonstrate the mathematical technique as based on the Langevin – Fokker-Planck – imaginary time Schrödinger scheme with subsequent symplectic integration. The specification of a physical system is given by energy functionals of growing complexity for quartic, sixth order and nonlocal Hamiltonians including those with electroelastic terms derived from first principles modeling. The test solutions reproduce effects of finite size, spatial inhomogeneity, ergodicity breaking, and alternate source field. Special attention is paid to bifurcation of stationary states and divergence of relaxation time captured by energy functionals modeling an assembly of coarse grained particles with attractive global and first neighbor coupling. Concluding, it is shown how the Langevin, Fokker-Planck and imaginary time Schrödinger equation techniques can be derived elegantly in terms of symplectic integration even for nonlocal and hardly nonlinear problems and can be used in calculations of response properties of spatially inhomogeneous metastable systems.

## Acknowledgements

This work has been partially supported by the Contract No. ICA1-CT-2000-70007 of European Excellence Center of Advanced Material Research and Latvian Science Project Nr.01.0805.1.1.